\documentstyle[prc,preprint,tighten,aps]{revtex}
\begin {document}
\draft

\title{Solution of few-body problems with the 
stochastic variational method: I. Central forces}

\author{K. Varga$^{1,2}$ and Y. Suzuki$^2$
\\
$^1$Institute of Nuclear Research, Hungarian Academy of Sciences,
(ATOMKI),
H-4001 Debrecen, Hungary
\\$^2$Department of Physics, Niigata University,
Niigata 950-21, Japan\\}
\maketitle
\date{\today}

\begin{abstract}
This paper presents a fortran program to solve diverse few-body problems 
with the stochastic variational method. Depending on the available
computational resources the program is applicable for 
$N=2-3-4-5-6-...$-body systems with $L=0$ total orbital momentum.
The solution with the stochastic variational method is ``automatic'' 
and universal. One defines the system (number of particles, masses, symmetry, 
interaction, etc. ) and the program finds the ground state energy and wave
function. The examples include nuclear (alpha particle: four-body, $^6$He:
six-body), atomic (td$\mu^-$ and $e^+e^-e^+e^-$) and subnuclear
(the nucleon and the delta in a nonrelativistic quark model)
systems. The solutions are accurate for excited states as well, and
even the Efimov-states can be studied. 
The program is available from the author (e-mail:varga@rikaxp.riken.go.jp).
\end{abstract}


\narrowtext

\section{Introduction}
\par\indent
  The solution of few-body problems is an important basic 
problem of physics.  One encounters few-body problems
from celestial to quark level in atomic, chemical, nuclear
or subnuclear physics. In recent years, due to the intense
experimental, theoretical and technological interest in 
mesoscopic scale systems in solid state physics (few ions in 
a trap, few electrons in a quantum dot, etc.)  the traditional 
domain of application is greatly enlarged.

\par\indent
  In the last few years we have elaborated a powerful 
technique, the stochastic variational method (SVM) 
\cite{VS,VSL,Kuk}, which is proved 
to be especially suitable for solution of diverse few-body 
problems. The stochastic variational method optimizes the 
variational basis in a random trial and error procedure. The
basis selection is free from any bias, keeps the dimension of
the basis low, and lastly but most importantly, provides a very
accurate solution. The method can be used with different type
of bases. The correlated Gaussian basis \cite{Szal} seems to be 
particularly suitable for description of $N=2-8$ particle systems, while for
larger number of particles displaced Gaussian or harmonic oscillator
bases can be applied.

\par\indent
  The method is a natural extension of the rigorous few-body ($N=3,4$)
techniques [5-11] to larger systems of strongly correlating particles and
offers a wide range of applications.

\par\indent
The aim of this paper is to present  a computer code for solution
of few-body problems with the stochastic variational method on correlated
Gaussian bases. The particles can interact via different central 
(Coulomb, Yukawa, Gauss or power law or other numerically given) 
potentials. The interaction may contain spin-isospin dependent 
operators. 
The program is general: the 
number of particles is in principle arbitrary, in practice it is limited by 
the speed and the memory of the available computer. The method can be used
to get very accurate solution for smaller systems or to find an approximate
upper bound for larger systems. One can treat bosons and fermions.

\par\indent
The applicability of the program is shown with various examples, including 
nuclear, Coulombic and quark systems. The accuracy of the solutions are
tested by comparing the results to those of the literature. An example
of the Efimov-states \cite{efi} shows that the method gives precise 
energies for ground and excited states.

\par\indent
The plan of the paper is as follows. In section 2 we outline the 
method.
In section 3 we show the calculation of the matrix elements. The 
fortran code is described in section 4. Examples are presented 
in section 5. The papers ends with a brief summary.

\section{The stochastic variational method}

Let's consider an $N$-particle system, where 
the $i$th particle with
mass $m_i$, spin $s_i$, isospin $t_i$ and charge $z_i$ is placed at the 
position ${\bf r}_i$.
The positions of the particles can be more conveniently described by 
introducing
a set of relative (Jacobi) coordinates ${\bf x}=({\bf x}_1,...,{\bf x}_{N-1})$ 
and the coordinate of the centre of mass ${\bf x}_N$.
The object of this paper is to solve the many-body Schrodinger-equation

\begin{equation}
{\cal H}\Psi = E \Psi \ \ \ \ \ \ 
({\cal H}=\sum_{i=1}^{N} {{\bf p}_i^2 \over 2 m_i}-T_{cm} +\sum_{i<j}^{N} 
V_{ij}) 
\end{equation}

\par\noindent
for central two-body interactions $V_{ij}$.
\par\indent
In our variational approach the basis functions are assumed to 
have the form

\begin{equation}
\psi_{SM_STM_T}({\bf x},A)={\cal A}\lbrace 
G_A({\bf x})\chi_{SM_S} \eta_{TM_T}
\rbrace,
\ \ \ \ \ \ 
G_A({\bf x})={\rm e}^{-{1\over 2} {\bf x} A {\bf x}} ,
\end{equation}

\par\noindent
where the operator ${\cal A}$ is an antisymmetrizer,
$\chi_{SM_S}$  is the spin function, and $\eta_{TM_T}$  is the 
isospin function of the system (for Coulombic system this latter 
can be suppressed). The diagonal elements of the $(N-1)\times (N-1)$ 
dimensional symmetric, positive definite matrix $A$ corresponds 
to the nonlinear parameters of an Gaussian expansion, and 
the off diagonal elements connect the different relative 
coordinates representing the correlations between the particles. 
This trial function, the correlated Gaussian basis, is widely 
used in physics \cite{Szal,KA91,KP93}, although the applications are mostly 
restricted to three and four particle systems. The above form assumes orbital 
angular momentum $L=0$.

\par\indent
In the variational method the wave function of the system is expanded as

\begin{equation}
\Psi=\sum_{i=1}^{K} c_i \psi_{SM_STM_T}({\bf x},A_i),
\end{equation}

\par\noindent
and an upper bound for the ground state energy of the system is 
given by the lowest eigenvalue of the generalized eigenvalue problem

\begin{equation}
H C = E_K B C,
\end{equation}

\par\noindent
where

\begin{equation}
H_{ij}=\langle \psi_{SM_STM_T}({\bf x},A_i)\vert {\cal H}\vert
\psi_{SM_STM_T}({\bf x},A_j)\rangle \ \ \ \ \ \  ,
B_{ij}=\langle \psi_{SM_STM_T}({\bf x},A_i)\vert
\psi_{SM_STM_T}({\bf x},A_j)\rangle \ \ \ \ \ \ 
\end{equation}

\par\noindent
The adequate choice of the nonlinear parameters (the elements of the 
$(N-1)\times(N-1)$ matrix $A$) is crucial. If these parameters are 
properly optimized, the variational solution gives very precise
energies. Although several ways for the choice of elements of
$A$ have been suggested, there is no safe recipe available. While 
the numerical 
optimization would be in principle the method of choice, in practice
there are several difficulties to face. The most serious ones amongst these 
are probably the large number of parameters to be optimized  and the 
nonorthogonality of the basis states. Due to the nonorthogonality, none of the
parameter sets is indispensable, and several different choices can represent
the wave functions equally well. This property makes the optimization 
tedious but offers the possibility of  random selection of the nonlinear 
parameters.

\par\indent
In the stochastic variational method the most suitable parameters are 
selected as follows:

\par\noindent
(1) several sets of $(A_1^n,....,A_K^n, n=1,...,{\cal N})$ are generated 
randomly,

\par\noindent
(2) by solving the eigenvalue problem the corresponding energies 
$(E_K^{1},...E_K^{\cal N})$ are determined,

\par\noindent
(3) the parameter set $(A_1^n,....,A_K^n)$ belonging to the 
lowest energy $E_K^{n}$  are chosen as basis parameters.

\par\indent
The bottleneck of the above procedure is that it assumes a 
full diagonalization to solve the eigenvalue problem and 
beyond a certain dimension it becomes too computer time
consuming. In practical implementation it is more advantageous
to fix $(A_1,....,A_{K-1})$ and change only $A_K^n$. In this
case only one row (column) of the matrices $H$ and $B$ has been
changed, and no diagonalization is needed but one can utilize the 
formulae of ref. [3]. Moreover, as by increasing the number 
of basis states the energy decreases ($E_{K+1}\le E_{K}$), we found it 
advantageous to increase the dimension of the basis after selecting 
the most suitable $A_K^n$ amongst the random candidates. This step ensures
that the energy $E_K^{n}$, belonging to a new, randomly selected $(A_K^n)$
basis state, is lower than the energy $E_{K-1}$ on the previously 
found basis, and provides a convenient selection criteria.
The steps
of the most economical way is therefore:

\par\noindent
(i) several sets of $(A_K^n, n=1,...,{\cal N})$ are generated 
randomly,

\par\noindent
(ii) by solving the eigenvalue problem the corresponding energies 
$(E_K^{1},...E_K^{\cal N})$ are determined,

\par\noindent
(iii) the parameter  $(A_K^n)$ belonging to the 
lowest energy $E_K^{n}$  chosen as a basis parameter and
by adding it to the previous basis states the parameters of
the new basis become  $(A_1,....,A_{K-1},A_K=A_K^n)$,

\par\noindent
(iv) the basis dimension is increased to $K+1$.

\section{Matrix elements}

One of the main advantages of the correlated gaussian basis is
that its matrix elements can be easily calculated analytically.
We list these matrix elements in the following.

\par\indent
The overlap of the correlated Gaussians takes the simple form:

\begin{equation}
\langle G_A\vert G_{A'}\rangle=
\left({(2\pi)^{(N-1)} \over {\rm det}(A+A')}\right)^{3 \over 2}.
\end{equation}

\par\noindent
The matrix element of the kinetic energy between the correlated Gaussians 
reads as:

\begin{equation}
\langle G_A\vert 
\sum_{i=1}^{N}{{\bf p}_i \over 2m_i}-T_{cm}
\vert G_{A'}\rangle=
\langle G_A\vert G_{A'}\rangle
\left(3{\rm Tr}(\Lambda A)-3{\rm Tr}(A+A')^{-1}(A'\Lambda A')\right),
\end{equation}

\par\noindent
where $\Lambda$ is an $(N-1)\times(N-1)$ diagonal matrix

\begin{eqnarray}
\Lambda=
\left(
\begin{array}{cccc}
 \hbar^2\over2\mu_1 & 0 & ...&  0       \\
 0  & \hbar^2\over2\mu_2 &    &  \vdots    \\
\vdots   &       &      &  \vdots         \\ 
 0  &...  & ...  & \hbar^2\over2\mu_{N-1} 
\end{array}
\right),
\end{eqnarray}

\par\noindent
and the reduced masses are given by

\begin{equation}
\mu_i={m_{i+1}m_{12\cdot\cdot\cdot{i}} \over m_{12\cdot\cdot\cdot{i+1}}} 
 \ \ \ \   (i=1,...,N\!\!-\!\!1), 
\ \ \ \ \ \ \ {\rm and} 
\ \ \ \ \ \mu_N=m_{12\cdot\cdot\cdot{N}},
\end{equation}

\par\noindent
with $m_{12\cdot\cdot\cdot{i}}=m_1\!+\!m_2\!+\cdot\cdot\cdot+\!m_i$.

\par\indent
To avoid the dependence of the matrix elements of the two-body interaction
on the specific form of the potential, it is advantageous to express the
potential in the form

\begin{equation}
V({\bf r}_{i}-{\bf r}_j)=\int d{\bf r} V({\bf r})
\delta({\bf r}_i -{\bf r}_j-{\bf r}).
\end{equation}

\par\noindent
To calculate the matrix elements of the potential we first express 
the relative distance vector
${\bf r}_i-{\bf r}_j$ 
by the Jacobi coordinates. The Jacobi 
coordinates
are defined as

\begin{equation}
{\bf x}_i=\sum_{j=1}^{N} U_{ij} {\bf r}_j,
\end{equation}

\par\noindent
with

\begin{equation}
U=
\left(
\begin{array}{ccccc}
-1            &     1     &    0 &... &  0              \\
-{m_1 \over m_{12}}  & -{m_2 \over m_{12}} &    1 &... &  0          \\
\vdots          &                &      &    &\vdots         \\ 
-{m_1 \over m_{12\cdot\cdot\cdot{N-1}}} &-{m_2 \over 
m_{12\cdot\cdot\cdot{N-1}}} 
& ...  &... & 1             \\
{m_1 \over m_{12\cdot\cdot\cdot{N}}} & {m_2\over m_{12\cdot\cdot\cdot{N}}}   
 & ...  &... &{m_N \over m_{12\cdot\cdot\cdot{N}}} 
\end{array}
\right),
\end{equation}

\par\noindent
The relative distance vector between two particles then can be written as

\begin{equation}
{\bf r}_i-{\bf r}_j=\sum_{i=1}^{N-1}B_{ijk}{\bf x}_k, \ \ \ \ \ \ 
B_{ijk}=U_{ik}^{-1}-U_{jk}^{-1}.
\end{equation}

\par\noindent
By using this expression the matrix element of the potential is given by:

\begin{equation}
\langle G_A\vert 
\delta({\bf r}_i-{\bf r}_j -{\bf r})
\vert G_{A'}\rangle=
\langle G_A\vert G_{A'}\rangle
\left({1\over 2\pi p_{ij}}\right)^{3\over2}
{\rm e}^{-{r^2 \over 2 p_{ij}}}
\end{equation}

\par\noindent
where

\begin{equation}
p_{ij}=\sum_{k=1}^{N-1}\sum_{l=1}^{N-1}B_{ijk}(A+A')^{-1}_{kl}B_{ijl}.
\end{equation}

\par\indent
By integrating eq. (14)  over ${\bf r}$ one can recover the norm of the wave
function.

\par\noindent
To calculate the matrix element of the potential one has to multiply eq. (14)
by the radial form $V({\bf r})$ and integrate over ${\bf r}$:

\begin{equation}
\langle G_A\vert V_{ij}
\vert G_{A'}\rangle=
\int d{\bf r} V({\bf r})
\langle G_A\vert 
\delta({\bf r}_i-{\bf r}_j -{\bf r})
\vert G_{A'}\rangle=
\langle G_A\vert G_{A'}\rangle v(p_{ij}),
\end{equation}

\par\noindent
where

\begin{equation}
v(p_{ij})=
\left({1\over 2\pi p_{ij}}\right)^{3\over2}
\int d{\bf r} V({\bf r})
{\rm e}^{-{r^2 \over 2 p_{ij}}}
\end{equation}

\par\noindent
In the applications the two-body interaction is expressed as

\begin{equation}
V_{ij}=\sum_{i=1}^{N_o}\left( \sum_{k=1}^{N_t} V_k^i(\vert {\bf r}_i- 
{\bf r}_j\vert)\right) {\cal O}_i
\end{equation}
where the operators (${\cal O}_1,\ldots ,{\cal O}_4$) are the 
Wigner, Majorana, Bartlett and Heisenberg operators.

\par\indent
If the radial part can be given in the form

\begin{equation}
V_k^i(r)=r^n{\rm e}^{-ar^2+br}  (n\ge -2),
\end{equation}

\par\noindent
then the integration can be calculated analytically by using the 
formula

\begin{equation}
\int_0^{\infty} dr r^n {\rm}e^{-ar^2+br}={1\over 2} (-1)^n \sum_{k=0}^{n}
{n! \over (n-k)! k!} f(k) g(n-k),
\end{equation}
\begin{equation}
f(k)=\left({1\over 2p}\right)^k \sum_{i=0}^{[k/2]} 
{k!\over i! (k-2i)!} q^{k-2i}p^i
\end{equation}

\begin{equation}
g(0)={\rm erfc}(y), \ \ \ \ g(k)=(-1)^k {2\over {\sqrt{\pi}}}
\left({1\over 2{\sqrt{p}}}\right)^k H_{k-1}(y), \ \ \ \ (k>1), \ \ \ \
y={q \over 2 {\sqrt{p}}}.
\end{equation}

\par\noindent
In other cases one has to rely on a numerical integration. To calculate the
matrix elements, the function $v(p)$ has to be evaluated many times for many
different values of $p$. Both the analytical and the numerical evaluation
takes some time on the computer. This part of the computation can be 
made faster,
however, noticing that $v(p)$ is a rather simple smooth function of $p$ and it
can be easily interpolated. To this end, for a given potential we 
tabulate $v(p)$ 
at certain representative values of $p$ and during the computation of the 
matrix
elements $v(p)$ is interpolated in the necessary points. The precision of the
interpolation can be easily controlled.

\section{Symmetrization}

The antisymmetrizer ${\cal A}$ is defined as

\begin{equation}
{\cal A}=\sum_{i=1}^{N!} {p_i} {\cal P}_i,
\end{equation}

\par\noindent
where the operator ${\cal P}_i$ changes the particle indices
according to the permutation $(p_{1}^i,...p_{N}^i)$ of the numbers
$(1,2,...,N)$, and $p_i$ is the parity of that permutation. The effect
of this operator on the set of position vectors $({\bf r}_1,...,{\bf r}_N)$
is

\begin{equation}
{\cal P}_i ({\bf r}_1,...,{\bf r}_N)=({\bf r}_{p_1^i},...,{\bf r}_{p_N^i})
\end{equation}

\par\noindent
By representing the permutations by the matrix

\begin{equation}
\left(C_i\right)_{kj}=1 \ \ \ \  {\rm if} \ \ \ \  j=p_{k}^i \ \ \ \ {\rm and}
\ \ \ \ \left(C_i\right)_{kj}=0 \ \ \ \ {\rm otherwise},
\end{equation}

\par\noindent
(for example, if $N=3$  the permutation $(3\ 1\ 2 )$  
is represented by

\begin{eqnarray}
C=
\left(
\begin{array}{ccc}
0 & 0 & 1\\
1 & 0 & 0\\
0 & 1 & 0\\
\end{array}
\right),
\end{eqnarray}

\par\noindent
while for $(1\ 2\ 3 )$ $C$ is a unit matrix),
the effect of the permutation operator on the single particle coordinates

\begin{equation}
{\cal P}_i ({\bf r}_1,...,{\bf r}_N)=C_i({\bf r}_{1},...,{\bf r}_{N}).
\end{equation}

\par\noindent
By using  eqs. (11) and (27) the permutation of the relative coordinates is
expressible as

\begin{equation}
{\cal P}_i {\bf x}=P_i{\bf x},
\end{equation}

\par\noindent
where $P_i$ is an $(N-1)\times(N-1)$ matrix obtained by omitting the 
last row and column (corresponding to the permutation invariant 
center-of-mass coordinate) of the $N\times N$ matrix $U^{T}C_iU$.

\par\indent
The correlated Gaussian function, after permutation takes the form:

\begin{equation}
{\cal P}_i G_{A}({\bf x})=G_{P_i^TAP_i}({\bf x}).
\end{equation}

\par\noindent
This simple transformation property under permutation is particularly
useful in calculating the matrix elements of the antisymmetized 
basis functions.

\par\indent
In the spin-isospin space the permutation operator interchanges the indeces
the single particle spin-isospin functions and can be easily evaluated.

\section{Program description}

\subsection{Fortran code}

To prepare the data for the calculation, the main program calls the following 
subroutines:

\par\noindent
{\bf input\_data}: This subroutine reads the input files described in the
previous subsection.

\par\noindent
{\bf cm\_rel\_tr}: This subroutine is responsible for the separation of the
relative and centre-of-mass coordinates.

\par\noindent
{\bf me\_spiso}: The overlap of the spin-isospin part, and the $P_i$ matrices
are calculated in this segment.

\par\indent
The matrix elements of the kinetic energy and potential energy 
are calculated in the
subroutines {\bf vkin\_ene} and {\bf vpot\_ene} by using the formulae
(7) and (16). The overlap of the basis states, given by eq. (6), is
computed in {\bf vove\_mat}.
The subroutine {\bf mat\_elem} prepares the matrices $H$ and $B$ of 
eq. (4).

\par\indent
The parameters of the basis states can be selected in three ways.
The first (s1) is a direct calculation on a given basis, the second (s2) and 
third (s3) are two versions of the stochastic selection of the basis states.

\vskip 0.25cm
\par\noindent
(s1) One can use predefined basis. In this case, the parameters of the
basis has to be written in the file ``fbs.res''. The first line
of this file defines the number of basis states $K$. In the next lines
the parameters of the basis are listed. Each line contains the corresponding
nonlinear parameters $(A_{k})_{ij} \ \ \  (k=1,...,K)$.

\par\noindent
(s2) One can select the parameters randomly and set up the basis step
by step increasing the dimension, as described by (i)-(ii)-(iii)-(iv). 
This is a rather automatic procedure, the only thing one has to define
before starting is the interval from which the uniformly distributed
random numbers are generated.

\par\noindent
(s3) One can select the basis states through the steps (i)-(ii)-(iii),
on a fixed basis dimension $K$. Before starting this procedure, one has to
give initial values of the nonlinear parameters. Then a basis state, say
the last one (the $K$th) is subjected to the procedure (i)-(ii)-(iii). If
a better new parameter set is found, then the $K$th state is substituted
by the new parameters. This procedure is repeated several times cyclically
for the other elements of the basis as well. We refer to this step as 
refinement because in this way we can further improve a basis. Particularly,
the bases, selected in the (s1) or the (s2) manner can be refined with this
method.

\vskip 0.25cm
\par\indent
The subroutines {\bf preset, svm1, svm2} corresponds to the three
possibilities above. To select amongst the three possibilities 
the value of {\sl ico} should be set to 1,2 or 3.

\par\indent
One does not need to solve the whole eigenvalue problem in each time 
when a basis parameter is changed. The program utilizes the simplifications 
described in ref. [3].

\par\indent
The function ``$v(p)$'' in eq. (17) can be calculated in three ways.

\vskip 0.25cm
\par\noindent
(p1) One can numerically integrate over the radial coordinate in eq.
(17) for given values of $p$ and tabulate the function $v(p)$. 
This tabulation can be done before the real calculation is started.
During
the basis selection, one needs the value of $v(p)$ at arbitrary values
of $p$ and these  values will be approximated by interpolation.

\par\noindent
(p2) If the potential is linear combination of terms like in eq. (18)
one can use the analytical formula (19).

\par\noindent
(p3) In this case we numerically integrate over the radial variable
in eq. (17) for each value of $p$ that appear during the basis selection.
The difference between this case and the first one is that as the actual
value of $p$ is available only during the basis selection process, 
this numerical integration can only be done then.

\vskip 0.25cm
\par\indent
The order of the above three possibilities reflects their speeds. The first
way is the fastest by far, but it accumulates the inaccuracies of the
numerical integration and interpolation. In most of the practical cases
these inaccuracies can be kept under control. The second case is slower
but it is exact for Coulomb, linear, harmonic oscillator, Gaussian and
Yukawa potentials. The third case is quite slow. The recommended way
is to use the first possibility for the basis selection. Once the selection
is finished, the chosen basis can be considered as a predefined basis
and one can rerun the calculation on that basis recalculating the matrix
elements by the (p1) or (p2) method. The accuracy of the numerical integration
is controlled by the variable $eps$. To check the accuracy of the numerical
integration one may rerun the calculation for different values of $eps$.

\par\indent
To select amongst (p1)-(p2)-(p3) one has to create a file ``pot.inp''
and write a number ``ipcon'' (ipcon=1,2 or 3) into its first line. In 
the case (p2) one
should write the parameters of the potentials into the file ``pot.inp''
as well. The number of the linear combinations of terms like eq. (17) should
be written into the second line. It is to be followed by the coefficient of
the linear combination, and the parameters $a, b, n$ of the terms in the 
successive lines, in turn. In the cases (p1) and (p3), the radial form 
of the  potential  should be defined in the subroutine
``pot'' by the user. The value of ``pot'' should be equal to the value
of the two-body potential at the radial distance ``$x$''.

\subsection{Selection of nonlinear parameters}
The Jacobi coordinates are convenient, because the kinetic energy can 
be expressed simply (eq. (1)). The choice
of the nonlinear parameters, however, is simpler in a coordinate
system, where the interparticle distances ${\bf r}_i-{\bf r}_j)$ are used.
The basis function in that system takes the form:

\begin{equation}
{\rm exp}\lbrace -\sum_{i<j} \alpha_{ij}({\bf r}_i-{\bf r}_j)^2\rbrace .
\end{equation}

The elements of $A$ in $G_A({\bf x})$ can be expressed by $\alpha_{ij}$
using an appropriate linear transformation. The advantage of $\alpha_{ij}$
is that it is more directly connected to the interparticle distances and
it can be more uniformly used. In the practical applications $\alpha_{ij}$
are generated as random numbers from the $[b_{min},b_{max}]$ interval.
The elements of $A$ are then expressed by using the values of $\alpha_{ij}$.

\par\indent
Although one can choose the elements of $A$ (through  $\alpha_{ij}$)
independently of each other randomly, we have found that it is slightly
more advantageous to follow the following way. At first, one generates
random values for each element. Then the first element of the matrix changed
randomly $K_0$ times. The best (giving the lowest ground state energy)
parameter is selected and fixed. The second element is fixed in the same
way and this process is continued till the last element. This procedure is then
repeated $M_0$ times. This selection requires $N_0\times M_0$ evaluations.
Only one element of the matrix of the nonlinear parameters is changed
in each step and that leads to the possibility of fast evaluation.

\subsection{Input data}
The input data specifies the system and the interaction. One has to
define the number of particles, the spins, isospins, charges and
masses. The spin and isospin states are  represented by integers,
the 'down' state is coded as 1 and the 'up' state is coded as 2.
The input data is read from the input file ``fbs.inp''. The first
line defines the number of particles $N$, the second line lists
the masses of particles ($m_1,\ldots ,m_N$), the third line gives
the charges ($z_1,\ldots ,z_N$), the fourth
the isospins ($t_1,\ldots ,t_N$). The spin part of the system 
can be given as linear combination of spin states. The next line 
contains the number of linear combinations, and the succeeding lines
give the linear combination coefficients and the spin states.
The following line should contain the value of 
$\hbar^2/m$, an initial (negative integer) number for the random number 
generator and the control parameters {\sl ico} and {\sl ibf}. The first
parameter selects the method of basis optimization (s1),(s2) or (s3), while 
the second defines the symmetry of identical particles ({\sl ibf}=1 for 
fermions and {\sl ibf}=2 for bosons). The next line gives $M_0$, $K_0$
and $K$. The program generates random numbers for a given nonlinear
parameter, say $A_{11}$ $K_0$ times, and once the best parameters for
all $A_{ij}$ has been found this random selection is repeated $K_0$ 
times. 
\par\indent 
In the fortran program, the parameter ``mnbas'' and ''mnpar'' defines the 
maximum number of basis states and the maximum number of particles, 
respectively.

\subsection{Output data}

The main results of the calculation are the energies of the
ground and excited states and the wave function. The energies are 
written in the file ``ener.dat''. The basis dimension, the coefficients
of the wave function of the ground state and the nonlinear parameters
of the basis are in the file ``fbs.res'' after the calculation is 
finished.

\section{Examples}

In this part we show a few examples for the application of the program.
The computational time varies for different cases and strongly depends
on the basis size and on the number of particles.

\subsection{The $td\mu^-$ and the positronium molecule}
This example includes a three- and a four-body system with Coulomb
interaction. The $td\mu^-$ molecule is a bound system of two positively 
and a negatively charged particles with unequal masses. This system attracted
considerable attention in connection with the muon catalysed fusion 
\cite{kamimura}. The input parameters (the input files) are shown in Table I.
Atomic units are used. The isospin quantum number is not necessary in these
systems so it just stands for distinguishing of the particles with different
charges. 
The matrix elements of the Coulomb-potential can be analytically 
calculated and thus the second way (p2) is used.

\par\indent
One can start the calculation by a step-by-step random selection
of the basis states (s2-type, {\sl ico}=2). The first four digits of the 
ground state energy can be reproduced on a basis size of $K=50$ 
(see Table II.). To improve the energy on this basis 
size further, one is to restart the calculation with a refinement
circle (s3-type, {\sl ico}=3). By repeating the refinement three
times, the fifth digits of accuracy can be reached. 
By increasing the basis size more accurate result can be
found in need.
The energies in the tables are written into 
the output file ''ener.dat''.

\par\indent
One can solve the two-body problem as well. The above example
can be recalculated for two-body case ($t\mu^-$), by changing
the number of particles to 2 in the input file. One immediately
arrives at the energy of the $(t\mu^-)$ threshold (-99.64 a.u.).

\par\indent
The next example is the positronium molecule, the $e^+e^-e^+e^-$
Coulomb four-body system. This system has been subject of quite a few
study, but accurate calculations have become available only recently
\cite{VS,KA91,KP93}. In this case we have four particles with equal
masses and a nonadiabatic treatment is necessary. The spins of the 
particles are coupled to $S=0$. This spin coupling leads to a more symmetric
system and the convergency is much faster. Due to the mass
difference between the particles in this and in the previous case, the 
interval of the random numbers are to be chosen differently. The input files
are in Table III. The basis
selection follows the same way as in the previous example and the results
are compared in Table IV. to those of other calculations.
To increase the accuracy further one has to use a larger basis.
It is remarkable that this calculation, after refining cycles, gives
very close results on a basis dimension of $K=50$ to that of refs.
[1,13,14] with $K=300$ Correlated Gaussians. 
\par\indent
To see the tremendous effect of the spin coupling on the convergency
one may try to rerun the same calculation with using only one of the spin
configurations.

\subsection{The alpha particle and the $^6$He}
In this subsection we describe the application of the program
for nuclear systems with a simple central interaction. A four- and
a six-body system (the alpha particle and $^6$He) are chosen as example.

\par\indent
In the first example of this subsection the ground state energy of
the alpha particle is calculated with the central, spin-isospin independent 
Malfleit-Tjon V \cite{MTV} (MTV) interaction. The input is given in Table V. 
The spin coupling is the
same as in the case of the positronium molecule. Although the potential
matrix elements are analytical as before, the first
representation (p1) is suggested as an illustrative example. To this end 
one has to write a
function (in fortran) ``pot(r)'' which defines the MTV potential.
The program runs faster with this choice and the accuracy is still
appropriate for nuclear systems. The results are compared to other
calculations in Table VI. 

\par\indent
The next example is a six-body system. Usually only the Quantum
Monte Carlo [5,10] methods are capable of going beyond four particles.
In this example we try to solve the $^6$He with the central, spin-isospin
dependent Minnesota potential \cite{Min}. The potential, though again 
analytical, is used in
numerical form as a fortran function. To illustrate another option,
this example uses a predefined basis ({\sl ico}=1, s1-type).
The predefined basis is to be included in the file ``fbs.res''.
(The program package contains this file under the name ``fbs1.res''.)

\subsection{The nucleon and the delta in a nonrelativistic quark
model}

\par\indent
In this example the SVM is used to solve the nonrelativistic three-quark
problem. The nonrelativistic three-quark model of baryons has a long 
history (see e.g. refs. [18-21]). Many of the properties of the 
baryons are quite successfully explained in this framework 
\cite{nrqm1,nrqm2,quark2}. In this section 
we solve the three-body Schr\"odinger with a one-gluon exchange potential
\cite{quark1} 
\begin{equation}
V_{ij}={1\over 2}\left(-{\kappa\over r_{ij}}+{r_{ij}\over a}-D
+V_0{{\rm e}^{-r_{ij}/r_0}\over r_{ij}}{\sigma}_i\cdot{\sigma}_j\right)
\end{equation}
for the nucleon and the delta. The parameters of the potential is taken from 
ref. [20]. The three-quark problem with this potential has been studied
in the framework of the Faddeev-equations in ref. [21]. We use the same 
parameters and compare our results to theirs. 
\par\indent
The potential has no dependence on color degrees
of freedom therefore the antisymmetric color part of the wave function can
be factorized and the spin-flavor-space part of the wave function
must be symmetric (the control parameter ``ibf'' should be set to 2).
The input files for the nucleon and the delta  differ in the 
spin- and isospin- (flavor) part (see Table ). The potential is given in
a numerical form (ipcon=1). The quark masses is taken to be 337 MeV like 
in ref. [21]. 
\par\indent
Due to the confining potential the basis size required for 
a reasonably accurate solution is very small.
After suitable modifications of the symmetrization of the wave function
the program can be used to calculate other baryons as well \cite{quark3}.

\subsection{The Efimov states}

The last example shows the application for Efimov states \cite{efi}. 
A three-body system with short range forces may have several bound states.
If the two-body subsystem have just one bound state whose energy is (close to)
zero, the three-particle can interact at long distances and an infinite number 
of bound state may appear in the three-body system.  
These ''Efimov states'' are extremely interesting from both experimental 
and theoretical points of view because of their distinct properties.
These states are very loosely bound and their wave functions extend far 
beyond those of normal states. By increasing the strength of the two-body
interactions these states disappear.
\par\indent
A  three-boson system is considered, for simplicity. In this case we have 
three spinless identical particles with a symmetrized spatial wave
function. The potential between the particles is taken as a the P\"osch-Teller 
interaction because this potential is analytically solvable for the two-body 
case, and therefore the two-body binding energy  can be easily set to zero. 
A spatially very extended basis is needed to represent the weakly bound 
states. In this case we use a predefined basis (s1). The predefined basis
is created by using only the diagonal elements of the matrix $A$. These 
diagonal elements are taken as geometric progression:
\begin{equation}
(A_{k})_{11}={1\over (a_0 q_0^{(i-1)})^2} \ \ \ \ 
(A_{k})_{22}={1\over (a_0 q_0^{(j-1)})^2} \ \ \ \
(i=1,...,n,\ \ j=1,...,n, \ \  k=(n-1)i+j) .
\end{equation}
This construction defines an $n\times n$ dimensional basis. The parameters
are chosen as $n=20, a_0=0.1, x_0=2.4$. (This predefined basis can be 
found in the file ``fbs2.res''.)
To try this example one has to use the potential function ''pot'' with 
the P\"osch-Teller potential 
\cite{sam}:
\begin{equation}
V(r)={625.972\over {\rm sinh}(1.586\ r)^2}-
{1251.943\over {\rm cosh}(1.586\ r)^2} .
\end{equation}
The result of this calculation is given in Table XI. One can find the first
three Efimov states with this basis size. The ratios of the energies of the
bound states follow the $E_{i+1}/E_{i}={\rm exp}\lbrace -2\pi\rbrace$ rule
\cite{efi}. By increasing the basis size
one can reveal more bound excited states. Note that the value of $a_0 x_0^n$
roughly corresponds to the spatial extension of the basis. The present basis
in this example goes out up to $a_0\  x_0^n=0.1\times 2.4^9=1674990$ (fm). 
This extension is necessary to get the third bound state.
As a comparison, to calculate the ground state energy ($-4.81$) it is enough 
to choose $n=7$ and the basis covers only the $[0,10]$ (fm) interval.
The root mean square radii of the ground state is about 1.5 fm, while that 
of the first excited state is about 6000 fm. These facts illustrate the 
tremendous spatial extension of the Efimov states. 

\section{Summary}
A fortran program is presented which solves the few-particle Schr\"odinger-
equation by using the stochastic variational method. The usefulness and 
applicability of the method is illustrated on various examples. The program 
can be used to solve diverse few-body problems. The stochastic variational 
method selects the most important basis states and keeps the basis size 
low even for six-body problems. 
The solution with the SVM is ``automatic'' 
and universal. One defines the system (number of particles, masses, symmetry, 
interaction, etc. ) and the program finds the ground state energy and wave
function. The refining steps [24,25] applied here increase the accuracy
further. The number of particles can be easily increased up to a certain
(computer dependent) limit. The program can be used without the SVM,
by using a predefined basis as well. 

\par\indent
This work was supported by OTKA grant No. T17298 (Hungary) and
by Grants-in-Aid for Scientific Research 
(No. 05243102 and No. 06640381) and for International Scientific 
Research (Joint Research) (No. 08044065) of the Ministry of Education, 
Science and Culture (Japan). K. V. gratefully acknowledges the support 
of the JSPS. We are thankful for the possibility of using the computer
facilities of RIKEN.

\begin{table}

\caption{Input files for the $td\mu$ system.}

\begin{tabular}{ll}
fbs.inp         & \\
\hline
3                                 &  number of particles             \\
206.786,5496.918,3670.481         &  masses of particles             \\
$-$1,1,1                          &  charges of particles            \\
1,2,2                             &  isospins                        \\
1                                 &  number of spin configurations   \\
1.,1,2,1                          &  coefficient,spins               \\
1.0,$-$14491,2,1                  &  $\hbar^2/m$,{\sl iran}, {\sl ico},
{\sl ibf} \\
5,100,50                          &  $M_0,K_o,K$                     \\ 
0.0000001,0.1                     &  $b_{min},b_{max}$               \\
\hline
pot.inp                           &                                  \\
\hline
2                                 &  {\sl ipcon}                     \\
1,1                               &  $N_o,N_t$                       \\
1,0.,0.,$-$1                      &  parameters of the potential      
\end{tabular}
\end{table}

\begin{table}

\caption{Results  for the $td\mu$ system. (In atomic units.) }

\begin{tabular}{lc}
Energy                         &  basis selection \\
\hline
$-$111.291   a.u.  ($K=50$)      &  (s2)-type                               \\
$-$111.342   a.u.  ($K=50$)      &  (s2)-type followed by an (s3)-type      \\
$-$111.360   a.u.  ($K=50$)      &  (s2)-type followed by 3 times (s3)-type \\
$-$111.36444 a.u. ($K=200$)      &  ref. [25]                                \\
$-$111.36451 a.u. ($K=1442$)    &  ref. [15] 
\end{tabular}
\end{table}

\begin{table}

\caption{Input files for the positronium molecule.}

\begin{tabular}{ll}
fbs.inp         & \\
\hline
4                                 &  number of particles           \\
1.,1.,1.,1.                       &  masses of particles           \\
1,1,$-$1,$-$1                     &  charges of particles          \\
1,1,2,2                           &  isospins                      \\
4                                 &  number of spin configurations \\
 1.,1,2,1,2                       &  coefficient,spins             \\
-1.,1,2,2,1                       &                                \\
-1.,2,1,1,2                       &                                \\
 1.,2,1,2,1                       &                                \\
1.0,$-$14491,2                    &  $\hbar^2/m$,{\sl iran}, {\sl ico},
{\sl ibf}\\
5,25,50                           &  $M_0,K_0,K$                   \\ 
0.1,15.                           &  $b_{min},b_{max}$             \\
\hline
pot.inp                           &                                \\
\hline
2                                 &  {\sl ipcon}                    \\
1,1                               &  $N_o,N_t$                      \\
1,0.,0.,$-$1                      &  parameters of the potential      
\end{tabular}

\end{table}

\begin{table}

\caption{Results  for the positronium molecule. (Atomic units are used.) }

\begin{tabular}{lc}
Energy                          &  basis selection \\
\hline
$-$0.51548  a.u.  ($K=50$)      &  (s2)-type                               \\
$-$0.51575  a.u.  ($K=50$)      &  (s2)-type followed by an (s3)-type      \\
$-$0.51586  a.u.  ($K=50$)      &  (s2)-type followed by 3 times (s3)-type \\
$-$0.51600  a.u.  ($K=300$)     &  refs. [1,13,14] 
\end{tabular}

\end{table}

\begin{table}

\caption{Input files for the alpha particle.}

\begin{tabular}{ll}
fbs.inp         & \\
\hline
4                                 &  number of particles           \\
1.,1.,1.,1.                       &  masses of particles           \\
1,1,1,1                           &  charges of particles          \\
1,1,2,2                           &  isospins                      \\
4                                 &  number of spin configurations \\
 1.,1,2,1,2                       &  coefficient,spins             \\
-1.,1,2,2,1                       &                                \\
-1.,2,1,1,2                       &                                \\
 1.,2,1,2,1                       &                                \\
41.47,$-$14491,2                  &  $\hbar^2/m$,{\sl iran}, {\sl ico},
{\sl ibf} \\
5,25,50                           &  $M_0,K_0,K$                   \\ 
0.1,15.                           &  $b_{min},b_{max}$             \\
\hline
pot.inp                           &                                \\
\hline
1                                 &  {\sl ipcon}                   \\
1,1                               &  $N_o,N_t$                     \\
0.,0.,0.,0                        &  parameters of the potential      
\end{tabular}

\end{table}

\begin{table}

\caption{Results  for the alpha particle. (In MeV.) }

\begin{tabular}{lc}
Energy                         &  basis selection \\
\hline
$-$31.30    ($K=100$)       &  (s2)-type                               \\
$-$31.32    ($K=100$)       &  (s2)-type followed by an (s3)-type      \\
$-$31.33    ($K=100$)       &  (s2)-type followed by 3 times (s3)-type \\
$-$31.36    ($K=150$)       &  ref. [3] 
\end{tabular}

\end{table}

\begin{table}

\caption{Input files for the $^6$He.}

\begin{tabular}{ll}
fbs.inp         & \\
\hline
6                                 &  number of particles           \\
1.,1.,1.,1.,1.,1.                 &  masses of particles           \\
1,1,1,1,1,1,                      &  charges of particles          \\
1,1,2,2,2,2                       &  isospins                      \\
1                                 &  number of spin configurations \\
1.,1,2,1,2,1,2                    &  coefficient,spins             \\
41.47,$-$14491,2                  &  $\hbar^2/m$,{\sl iran}, {\sl ico},
{\sl ibf} \\
5,25,50                           &  $M_0,K_0,K$                   \\ 
0.1,15.                           &  $b_{min},b_{max}$             \\
\hline
pot.inp                           &                                \\
\hline
1                                 & {\sl ipcon}                        \\
1,1                               &  $N_o,N_t$                      \\
0.,0.,0.,0                        &  parameters of the potential      
\end{tabular}

\end{table}

\begin{table}

\caption{Results  for the $^6$He. (In MeV.) }

\begin{tabular}{lc}
Energy                     &  basis selection \\
\hline
$-$30.00    ($K=100$)      &  (s1)-type                               \\
$-$30.07    ($K=300$)      &  ref. [3] \\
\end{tabular}
\end{table}
\begin{table}

\caption{Input files for the nucleon and the delta particle .}

\begin{tabular}{ll}
fbs.inp (nucleon)                 & \\
\hline
3                                 &  number of particles           \\
1.,1.,1.                          &  masses of particles           \\
1,1,1                             &  charges of particles          \\
1,2,2                             &  isospins                      \\
2                                 &  number of spin configurations \\
1.,1,2,1                          &  coefficient,spins             \\
-1.,1,1,2                         &                                \\
115.54,$-$14491,2,2               &  $\hbar^2/m$,{\sl iran}, {\sl ico},
 {\sl ibf} \\
5,25,50                           &  $M_0,K_0,K$                   \\ 
0.1,8.                            &  $b_{min},b_{max}$             \\
\hline
fbs.inp (delta, spin-isospin part only  &                       \\
\hline
1,1,1                             &  isospins                      \\
1                                 &  number of spin configurations \\
1.,1,1,1                          &  coefficient,spins             
\end{tabular}

\end{table}

\begin{table}

\caption{Results for nucleon and delta (in MeV)  (3*337 MeV is added).}

\begin{tabular}{lc}
Energy                         &  basis selection \\
\hline
Nucleon                    &                 \\
\hline
1021       ($K=10$)       &  (s2)-type                               \\
1024                      &  ref. [21] \\
\hline
delta                     &                 \\
\hline
1330       ($K=5$)        &  (s2)-type                               \\
1330                      &  ref. [21] \\
\end{tabular}

\end{table}

\begin{table}
\caption{Input files for the Efimov states.}

\begin{tabular}{ll}
fbs.inp (nucleon)                 & \\
\hline
3                                 &  number of particles           \\
1.,1.,1.                          &  masses of particles           \\
1,1,1                             &  charges of particles          \\
1,1,1                             &  isospins                      \\
1                                 &  number of spin configurations \\
1.,1,1,1                          &  coefficient,spins             \\
41.47$,-$14491,1,2                &  $\hbar^2/m$,{\sl iran}, {\sl ico}, 
{\sl ibf} \\
5,25,50                           &  $M_0,K_0,K$                   \\ 
0.1,15.                           &  $b_{min},b_{max}$             \\
\hline
pot.inp                           &                                \\
\hline
1                                 &  {\sl ipcon}                    \\
1,1                               &  $N_o,N_t$                      \\
0.,0.,0.,0                        &  parameters of the potential    \\
\hline
output: ener.dat                  &                                 \\
\hline
dimension                         & 400                             \\
energy(1)                         & $-4.81$                         \\
energy(2)                         & $-8.77\times 10^{-3} $          \\
energy(3)                         & $-1.50\times 10^{-5} $               
\end{tabular}

\end{table}


\begin{references}
\bibitem{Kuk} V.I.Kukulin and V. M. Krasnopol'sky, J. Phys. G {\bf 3}  
(1977) 795.
\bibitem{VSL} K. Varga, Y. Suzuki and R. G. Lovas, Nucl. Phys. {\bf A571} 
(1994) 447.
\bibitem{VS}  K. Varga and Y. Suzuki, Phys. Rev. C {\bf 52} (1995) 2885.
\bibitem{Szal} S. A. Alexander, H. J. Monkhorst and K. Szalevicz, J. Chem.
Phys. {\bf 85} (1986) 5821, and references therein.
\bibitem{GFMC1} J. Carlson, Phys. Rev. C {\bf 36} (1987) 2026.
\bibitem{CHH1} M. Viviani, A Kievsky and S. Rosati, Few-Body Systems {\bf 18} 
(1995) 25. 
\bibitem{Kam} H. Kameyama, M. Kamimura and Y. Fukushima, Phys. Rev. 
{\bf C40} (1989) 974.
\bibitem{FY1} W. Gl\"ockle and H. Kamada, Phys. Rev. Lett {\bf 71} (1993) 971
\bibitem{Friar} C. R. Chen, G. L. Payne, J. L. Friar and B. F. Gibson,
Phys. Rev. C {\bf 31} (1985) 2266.
\bibitem{A6} B. S. Pudliner, V. R. Pandharipande, J. Carlson and 
R. B. Wiringa,	Phys. Rev. Lett. {\bf 74} (1995) 4396.	
\bibitem{Aka} Y. Akaishi, Few-body Systems, Suppl. {\bf 1} (1987) 120.
\bibitem{efi} V. Efimov, Phys. Lett. B{\bf 33}, (1970) 563. 
\bibitem{KA91}  P. M. Kozlowski and L. Adamowicz, J. Chem. Phys. {\bf 95}, 
6681 (1991).
\bibitem{KP93} D. B. Kinghorn and R. D. Poshusta, Phys. Rev. 
A{\bf 47}, 3671 (1993).
\bibitem{kamimura} M. Kamimura, Phys. Rev. A{\bf 38}, 621 (1988). 
\bibitem{MTV} R. A. Malfliet and J. A. Tjon, Nucl. Phys. {\bf A127}, 
161 (1969).
\bibitem{Min} D. R. Thompson, M. LeMere, and Y. C. Tang, 
Nucl. Phys. {\bf A286}, 53 (1977); I. Reichstein and Y. C. Tang, Nucl. Phys. 
{\bf A158}, 529 (1970).
\bibitem{nrqm1} M. Oka and K. Yazaki, Phys. Lett. {\bf B90}, 41 (1980).
\bibitem{nrqm2} N. Isgur and G. Karl, Phs. Rev. D {\bf 18}, 4187 (1978).
\bibitem{quark1} R. K. Bhaduri, L. E. Cohler, and Y. Nogami, Nouvo
Cimento 65A, 376 (1981). 
\bibitem{quark2} B. Silvestre-Brac and C. Gignoux,
Phys. Rev. D32 (1985) 743.
\bibitem{quark3} L. Ya. Glozman, Z. Papp, W. Plessas, K. Varga, 
R. F. Wagenbrunn, Nucl. Phys. A (in press).
\bibitem{sam} The authors are grateful to S. A. Moszkowski for suggesting this
potential.
\bibitem{vos} K. Varga, Y. Ohbayashi, and Y. Suzuki, Phys. Lett. B, in press.
\bibitem{suv} J. Usukura, Y. Suzuki and K. Varga, to be published 
in Phys. Rev. A.

\end{references}
\end{document}